# A Study on the Situation of Connected Car Patent Portfolios

Abel C. H. Chen[1,*], Chia-Shen Chang[2]
[1]Information & Communications Security Laboratory, Chunghwa Telecom Laboratories
[2]ICT Integration Department, Chunghwa Telecom Laboratories

**Abstract**

In recent years, the countries of the world have drafted the specifications of connected cars; for instance, the Security Credential Management System (SCMS) has been proposed by United States Department of Transportation (USDOT), and the Cooperative Intelligent Transportation System (C-ITS) Credential Management System (CCMS) has been proposed by European Union (EU). Therefore, several companies have developed the technology and productions of connected cars based on specifications, and connected car patent portfolios have been proactively performed. Therefore, this study uses Patent Search System (PSS) to find and analyze the contents of patents for obtaining the innovation reports of connected cars according to patents. This study considers the single-factor and two-factors to analyze the relationships of annuals, major technology leaders, major market leaders, and major technology and applications for exploring the patent portfolios of technology leaders and market leaders in connected cars.

## 車聯網專利佈局現況研究

### 摘要

近年來，世界各國開始制訂符合各自國情的車聯網規範，如：美國交通部安全憑證管理系統(Security Credential Management System, SCMS)、歐盟協同智慧運輸系統憑證管理系統(Cooperative Intelligent Transportation System (C-ITS) Credential Management System, CCMS)。因此，許多企業開始在符合標準下研發車聯網技術和產品，並且積極進行車聯網專利佈局。有鑑於此，本研究為分析車聯網創新技術的發展趨勢，採用專利資訊檢索系統，並且在系統中進行專利檢索和分析專利內容。本研究分別從單變量和雙變量的方式，觀察年度、主要技術領導企業、主要市場領導企業、以及主要技術和應用領域之間的關聯，並且探討主要技術領導企業和主要市場領導企業在各個主要技術和應用領域的專利佈局。
關鍵詞：車聯網、專利佈局、主要技術領導企業、主要市場領導企業

## 壹、 引言和研究目標

車聯網的研究和標準已經在十幾年前就開始進行，包含電機電子工程師學會（Institute of Electrical and Electronics Engineers, IEEE）、歐洲電信標準協會（European Telecommunications Standards Institute, ETSI）、國際汽車工程學會（Society of Automotive Engineers International, SAE International）等標準組織已經陸續為車聯網領域制定通訊標準、資訊安全標準、訊息格式標準等[1]-[3]。然而，雖然已經許多標準制定好初稿和技術文件，但仍需各個國家政府介入規範要求車聯網相關終端設備和網路設備需符合哪些標準，才能讓車聯網相關企業依循和生產符合標準的終端設備和網路設備，最終才能打造出車聯網生態系統。以車聯網資訊安全憑證管理系統為例，近年來美國在 IEEE 標準為主體的基礎上提出安全憑證管理系統(Security Credential Management System, SCMS)[4]-[6]，歐盟也在 ETSI 標準為主體的基礎提出協同智慧運輸系統憑證管理系統(Cooperative Intelligent Transportation System (C-ITS) Credential Management System, CCMS)[7]-[8]，制定符合北美環境和歐盟環境的車聯網標準規範。

隨著世界各國政府陸續明確採用的車聯網標準，以及建立符合各自國家情況的車聯網標準規範，也帶動車聯網相關企業的投入和積極進行技術及市場佈局。有鑑於此，研究車聯網專利佈局及其發展趨勢是個重要的研究議題。本研究將以專利資訊檢索系統為例，檢索車聯網的已獲證發明專利，並且分析歷年發展概況變化、主要技術領導企業、主要市場領導企業、以及主要技術和應用領域等。本研究的研究問題條列如下：

(1). 車聯網的歷年發展概況變化如何？
(2). 車聯網的主要技術領導企業有哪些？
(3). 車聯網的主要市場領導企業有哪些？
(4). 車聯網的主要技術和應用領域有哪些？

本論文主要分為五個章節。第貳節將介紹專利檢索系統與搜尋策略，說明搜尋關鍵字設計和搜尋結果。第參節先以單變量的方式，考慮年度、第一申請人、第一專利權人、國際專利分類(International Patent Classification, IPC)號等變量，進行各別分析和討論，用以觀察主要技術領導企業、主要市場領導企業、以及主要技術和應用領域。第肆節將採用雙變量的方式，將變量兩兩組合，考慮年度與第一申請人、年度與第一專利權人、第一申請人與國際專利分類號、以及第一專利權人與國際專利分類號等組合，用以觀察主要技術領導企業和主要市場領導企業在各個主要技術和應用領域的專利佈局。第伍節將總結本研究的發現和限制，並且提供未來建議。

## 貳、 專利檢索系統與搜尋策略

為分析車聯網專利佈局現況和發展趨勢，本研究採用專利資訊檢索系統進行專利搜尋和分析。在本研究中，以「(車聯網 OR 車載 OR V2X OR V2I) AND (NOT 化合物)」作為搜尋策略，並且主要分析「發明專利」，如圖 1 所示[9]。其中，由於許多車聯網相關專利採用「V2X」和「V2I」等關鍵字，但化合物部分發明專利也包含這些關鍵字，所以在搜尋策略中過濾掉包

含「化合物」關鍵字的發明專利。由於本研究以專利權人資訊作為觀察市場領導企業的指標，所以本研究僅分析「已核准發明專利」，而審核中和未核准的發明專利不在本研究範圍內。由於經濟部智慧財產局於每個月的 01 日、11 日、以及 21 日更新資料庫和系統，所以本研究的檢索日期是 2022 年 11 月 02 日，可以取得 2022 年 11 月 01 日（含）前核准的發明專利。檢索結果顯示從 1989 年到 2022 年期間，總共有 4,401 件車聯網相關的發明專利被核准，並且本研究將在此搜尋結果上進行分析和討論。

**圖 1　專利搜尋結果[9]**

為分析專利的技術領導企業、市場領導企業、以及技術和應用領域，本研究分別以專利的、第一申請人、第一專利權人、以及第一國際專利分類號作為觀察指標，詳述如下：

(1). 在專利申請時，同一件專利可能有一到多個申請人，通常第一申請人對該專利的技術掌握程度較高，所以本研究以第一申請人作為觀察技術領導企業的指標。並且，為觀察主要技術領導企業，本研究將以核准發明專利數前十名的第一申請人作為主要技術領導企業的指標。

(2). 在專利被核准後，同一件專利可能有一到多個專利權人，假設專利鑑價後，通常第一專利權人對該專利的授權金可以擁有較高的比例，所以本研究以第一專利權人作為觀察市場領導企業的指標。並且，為觀察主要市場領導企業，本研究將以核准發明專利數前十名的第一專利權人作為主要市場領導企業的指標。

(3). 在專利審查的過程中，經濟部智慧財產局對同一件專利可能劃分一到多個國際專利分類號，通常第一國際專利分類號是最適合該專利的類別，所以本研究以第一國際專利分類號作為觀察技術和應用領域的指標。並且，為觀察主要技術和應用領域，本研究將以核准發明專利數前十名的第一國際專利分類號作為主要技術和應用領域的指標。

## 參、 單變量分析

本節採用單變量分析的方式，從年度、第一申請人、第一專利權人、第一國際專利分類

號等單變量來分析發技術領導企業、市場領導企業、以及主要技術和應用領域展現況。

一、 歷年發展概況變化情況

　　本研究採集 1989 年到 2022 年期間已核准的 4,401 件發明專利，依年度劃分結果如圖 2 和表 1 所示。根據本研究統計分析，4,401 件發明專利裡，申請到核准所需花費的時間，審核最久的專利花費 12 年，中位數和平均值分別為 2 年和 2.75 年。由此可知，仍有較多的專利需要 2 年以上的時間才會被核准。

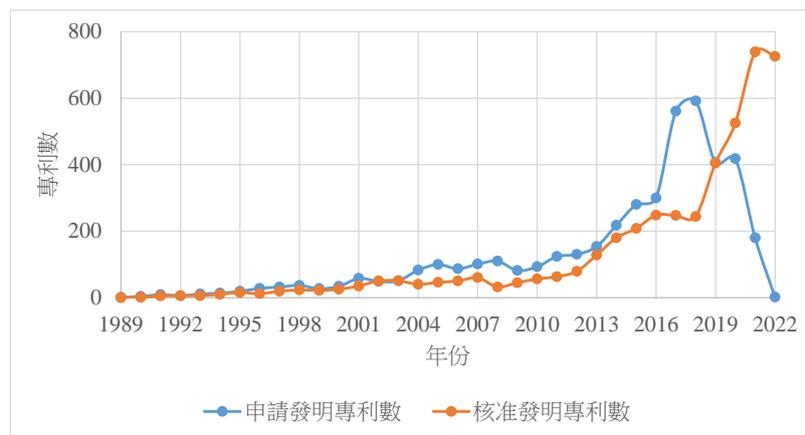

**圖 2　申請發明專利數和核准發明專利數的歷年發展概況變化情況**

　　由圖 2 呈現趨勢及第壹節所述觀察，車聯網技術已經行之有年，所以有部分企業在 1989 年開始佈局，並且隨著國際標準組織開始制定標準和引領學界及研究單位的投入研發，所以從 2000 年到 2016 年期間專利數有上升的趨勢，但增加速度不快。並且在 2016 年後，美國和歐盟陸續制定各自國家的車聯網標準規範，讓車聯網企業關注到政府未來可能會有明確的車聯網標準規範，所以開始加大投入的力道，申請發明專利數量快速增加；然而，由於審核核准時間平均是 2.75 年，所以 2017 年申請的發明專利陸續在 2019 年和 2020 年後被核准，所以核准發明專利數在 2019 年後快速增加。其中，因為 2019 年申請的專利到 2022 年 11 月時仍可能在審核中，所以申請發明專利數下降是仍有許多專利仍在審核中尚未核准所造成的。此外，從表 1 熱點圖表示方式中，2019 後每年核准發明專利數每年處於 400 件以上，是 2015 年之前每年核准發明專利數的數倍，反應出現在車聯網企業正積極投入專利佈局。

表 1  申請發明專利數和核准發明專利數的歷年發展概況變化情況

| 年份 | 申請發明專利數 | 核准發明專利數 |
|---|---|---|
| 1989 | 1 | 0 |
| 1990 | 4 | 1 |
| 1991 | 9 | 5 |
| 1992 | 6 | 6 |
| 1993 | 11 | 6 |
| 1994 | 14 | 10 |
| 1995 | 19 | 15 |
| 1996 | 28 | 12 |
| 1997 | 32 | 19 |
| 1998 | 37 | 23 |
| 1999 | 27 | 22 |
| 2000 | 34 | 25 |
| 2001 | 58 | 35 |
| 2002 | 49 | 50 |
| 2003 | 51 | 51 |
| 2004 | 83 | 40 |
| 2005 | 100 | 46 |
| 2006 | 87 | 50 |
| 2007 | 101 | 60 |
| 2008 | 110 | 32 |
| 2009 | 82 | 45 |
| 2010 | 93 | 56 |
| 2011 | 124 | 63 |
| 2012 | 130 | 79 |
| 2013 | 154 | 128 |
| 2014 | 218 | 180 |
| 2015 | 280 | 208 |
| 2016 | 300 | 248 |
| 2017 | 561 | 247 |
| 2018 | 592 | 244 |
| 2019 | 406 | 405 |
| 2020 | 418 | 525 |
| 2021 | 180 | 739 |
| 2022 | 2 | 726 |

## 二、 主要技術領導企業情況

本研究以第一申請人作為觀察技術領導企業的指標，並列舉核准發明專利數前十名的第一申請人作為主要技術領導企業，如表 2 所示。根據本研究統計分析，在檢索到的 4,401 件發明專利中，總共可以劃分到 1,137 個第一申請人，也就是平均每個第一申請人大約申請 3.87 件發明專利。然而，如果把核准發明專利數由高到低排序的話，會發現大部分專利集中在少數幾家企業，並且有 716 個第一申請人僅申請 1 件發明專利，呈現指數分佈。因此，觀察少數申請大量發明專利的企業（即主要技術領導企業）更有助於了解車聯網產業的發展情況。

表 2 核准發明專利數前十名的第一申請人（即主要技術領導企業）

| 第一申請人 | 核准發明專利數 |
| --- | --- |
| ＯＰＰＯ廣東移動通信有限公司 | **292** |
| 高通公司 | **244** |
| 新力股份有限公司 | **125** |
| 英業達股份有限公司 | **101** |
| 財團法人工業技術研究院 | **85** |
| 中華電信股份有限公司 | **67** |
| 鴻海精密工業股份有限公司 | **63** |
| 本田技研工業股份有限公司 | **63** |
| 聯發科技股份有限公司 | **57** |
| 北京市商湯科技開發有限公司 | **53** |

表 2 結果顯示，申請較多車聯網相關發明專利的技術領導企業是ＯＰＰＯ廣東移動通信有限公司和高通公司，分別核准了 292 件和 244 件。之後依序是新力股份有限公司、英業達股份有限公司、財團法人工業技術研究院、中華電信股份有限公司、鴻海精密工業股份有限公司、本田技研工業股份有限公司、聯發科技股份有限公司、以及北京市商湯科技開發有限公司。在後續章節中將對這些主要技術領導企業進行深入討論，分析其專利交互授權和專利佈局情況。

## 三、 主要市場領導企業情況

本研究以第一專利權人作為觀察市場領導企業的指標，並列舉核准發明專利數前十名的第一專利權人作為主要市場領導企業，如表 3 所示。根據本研究統計分析，在檢索到的 4,401 件發明專利中，總共可以劃分到 1,136 個第一專利權人，並且有 716 個第一專利權人僅擁有 1 件發明專利，呈現指數分佈，大致與第一申請人的情況相似。由於專利核准後，如果專利沒有做轉讓或授權等相關變動的情況下，第一申請人和第一專利權人會是相同的。因此，第一申請人和第一專利權人的分佈情況相似是一般現象。然而，本研究發現在少數申請大量發明專利的企業（即本研究定義的「主要技術領導企業」）和少數擁有大量發明專利的企業（即本研究定義的「主要市場領導企業」）不同，值得深入分析與討論。

表 3  核准發明專利數前十名的第一專利權人（即主要市場領導企業）

| 第一專利權人 | 核准發明專利數 |
|---|---|
| ＯＰＰＯ廣東移動通信有限公司 | 292 |
| 高通公司 | 244 |
| 新力股份有限公司 | 123 |
| 英業達股份有限公司 | 101 |
| 大唐移動通信設備有限公司 | 98 |
| 財團法人工業技術研究院 | 82 |
| 中華電信股份有限公司 | 67 |
| 本田技研工業股份有限公司 | 63 |
| 開曼群島商創新先進技術有限公司 | 57 |
| 鴻海精密工業股份有限公司 | 54 |

　　表 3 結果顯示，擁有較多車聯網相關發明專利的市場領導企業是ＯＰＰＯ廣東移動通信有限公司和高通公司，之後依序是新力股份有限公司、英業達股份有限公司、大唐移動通信設備有限公司、財團法人工業技術研究院、中華電信股份有限公司、本田技研工業股份有限公司、開曼群島商創新先進技術有限公司、以及鴻海精密工業股份有限公司。從核准發明專利數觀察，可以得知高通公司、英業達股份有限公司、中華電信股份有限公司、本田技研工業股份有限公司等企業主要都是自主研發的技術，並且主要用來保護公司自身的產品，未轉讓和未授權給其他企業。而其他第一專利權人的企業相較於該企業擔任第一申請人時的核准發明專利數則略有增加的情況，也就是有通過購買或其他方式取得發明專利的轉讓或授權。

　　其中，核准發明專利數增加較多的企業是大唐移動通信設備有限公司、開曼群島商創新先進技術有限公司這兩家公司，並且這兩家公司未在主要技術領導企業列表中，卻進入主要市場領導企業列表中，代表獲得較多專利的轉讓和授權，以下將深入討論。大唐移動通信設備有限公司其獲得核准發明專利的來源主要來自於其母公司電信科學技術研究院有限公司，而電信科學技術研究院有限公司申請獲得核准的發明專利大部分都轉讓或授權給大唐移動通信設備有限公司；因此，大唐移動通信設備有限公司的核准發明專利數從 36 件增加到 98 件。此外，開曼群島商創新先進技術有限公司的作法類似，其獲得核准發明專利的來源主要來自於其母公司阿里巴巴集團服務有限公司，阿里巴巴集團服務有限公司申請獲得核准的發明專利大部分都轉讓或授權給開曼群島商創新先進技術有限公司；因此，開曼群島商創新先進技術有限公司的核准發明專利數從 22 件增加到 57 件。

四、 主要技術和應用領域情況

　　本研究以第一國際專利分類號作為觀察技術和應用領域的指標，並列舉核准發明專利數前十名的第一國際專利分類號作為主要技術和應用領域，如表 4 所示。根據本研究統計分析，在檢索到的 4,401 件發明專利中，總共可以劃分到 249 個第一國際專利分類號，也就是平均每個第一國際專利分類大約申請 17.67 件發明專利。然而，如果把核准發明專利數由高到低

- 7 -

排序的話，會發現大部分專利集中在少數幾個國際專利分類號，核准發明專利數前十名的第一國際專利分類號其專利數總和有 2,309 件，即達到所有核准發明專利數的一半以上。因此，觀察少數具有大量發明專利的國際專利分類號更有助於了解車聯網產業的主要技術和應用領域情況。

表 4 結果顯示，具有較多車聯網相關發明專利的第一國際專利分類號是 H04W、H04L、以及 G06F，對應主要應用領域分別是無線通訊網路、數位資訊之傳輸、以及電子數位資料處理。其他的第一國際專利分類號和主要應用領域依序是 H04N（影像通信）、H01L（半導體裝置）、G06Q（數據處理系統或方法）、B60R（其他類不包括的車輛，車輛配件或車輛部件）、H04B（傳輸）、G02B（光學元件、系統或儀器）、G02F（用於控制光之強度、顏色、相位、偏振或方向之器件或裝置）[10]。

表 4  核准發明專利數前十名的第一 IPC（即主要技術和應用領域）[10]

| 第一 IPC | IPC 說明 | 核准發明專利數 |
|---|---|---|
| H04W | 無線通訊網路 | **640** |
| H04L | 數位資訊之傳輸 | **330** |
| G06F | 電子數位資料處理 | **304** |
| H04N | 影像通信 | **193** |
| H01L | 半導體裝置 | **187** |
| G06Q | 專門適用於行政、管理、商業、經營、監督或預測目的的數據處理系統或方法 | **160** |
| B60R | 其他類不包括的車輛，車輛配件或車輛部件 | **157** |
| H04B | 傳輸 | **120** |
| G02B | 光學元件、系統或儀器 | **117** |
| G02F | 用於控制光之強度、顏色、相位、偏振或方向之器件或裝置 | **101** |

由於車聯網大部分通過無線通訊網路進行傳輸，並且有少部分通過有線通訊網路進行傳輸，並著重在資料通訊和處理。因此，在核准發明專利中，有 622 件已核准發明專利劃分到 H04W，即無線通訊網路領域。並且，車聯網涉及資料傳輸和資料處理，所以分別有 330 件、304 件劃分到 H04L（數位資訊之傳輸）、G06F（電子數位資料處理）。在第肆節中，將以雙變量的方式對各個主要技術領導企業和各個主要市場領導企業在各個主要技術和應用領域的專利佈局進行深入討論。

## 肆、 雙變量分析

本節採用雙變量分析的方式，分別從主要技術領導企業的歷年發展概況變化、主要技術領導企業的主要技術和應用領域、以及主要市場領導企業的主要技術和應用領域等三個面向進行討論。

## 一、 主要技術領導企業的歷年發展概況變化情況

為分析主要技術領導企業的歷年發展概況變化情況，本研究從年度和第一申請人兩個變量來分析其關聯，並且以核准發明專利數前十名的第一申請人作為主要技術領導企業進行觀察和討論，如表 5 所示。

表 5 核准發明專利數前十名的第一申請人在各年度的核准發明專利數

| 年份 | OPPO廣東移動通信有限公司 | 高通公司 | 新力股份有限公司 | 英業達股份有限公司 | 財團法人工業技術研究院 | 中華電信股份有限公司 | 鴻海精密工業股份有限公司 | 本田技研工業股份有限公司 | 聯發科技股份有限公司 | 北京市商湯科技開發有限公司 |
|---|---|---|---|---|---|---|---|---|---|---|
| 2001及之前 | 0 | 0 | 0 | 0 | 1 | 0 | 0 | 9 | 0 | 0 |
| 2002 | 0 | 0 | 0 | 0 | 1 | 0 | 0 | 2 | 0 | 0 |
| 2003 | 0 | 0 | 0 | 1 | 0 | 1 | 1 | 7 | 0 | 0 |
| 2004 | 0 | 0 | 0 | 0 | 1 | 0 | 0 | 1 | 0 | 0 |
| 2005 | 0 | 0 | 1 | 0 | 0 | 2 | 0 | 4 | 0 | 0 |
| 2006 | 0 | 0 | 1 | 0 | 2 | 0 | 0 | 2 | 0 | 0 |
| 2007 | 0 | 0 | 0 | 0 | 4 | 1 | 1 | 2 | 0 | 0 |
| 2008 | 0 | 0 | 0 | 0 | 0 | 0 | 1 | 0 | 0 | 0 |
| 2009 | 0 | 0 | 2 | 0 | 1 | 1 | 2 | 1 | 1 | 0 |
| 2010 | 0 | 3 | 1 | 0 | 2 | 1 | 3 | 2 | 0 | 0 |
| 2011 | 0 | 1 | 5 | 0 | 1 | 1 | 7 | 1 | 1 | 0 |
| 2012 | 0 | 4 | 3 | 0 | 0 | 4 | 4 | 3 | 0 | 0 |
| 2013 | 0 | 13 | 3 | 0 | 3 | 2 | 5 | 0 | 2 | 0 |
| 2014 | 0 | 4 | 5 | 0 | 9 | 6 | 5 | 4 | 1 | 0 |
| 2015 | 0 | 7 | 1 | 1 | 7 | 2 | 7 | 8 | 0 | 0 |
| 2016 | 0 | 17 | 2 | 1 | 9 | 4 | 8 | 6 | 0 | 0 |
| 2017 | 0 | 4 | 2 | 1 | 5 | 9 | 4 | 0 | 0 | 0 |
| 2018 | 0 | 6 | 11 | 0 | 8 | 10 | 2 | 1 | 2 | 0 |
| 2019 | 5 | 14 | 22 | 1 | 15 | 7 | 4 | 1 | 4 | 0 |
| 2020 | 41 | 39 | 36 | 1 | 5 | 8 | 2 | 7 | 14 | 2 |
| 2021 | 108 | 57 | 23 | 34 | 8 | 4 | 1 | 1 | 25 | 14 |
| 2022 | 113 | 75 | 7 | 61 | 3 | 4 | 6 | 1 | 6 | 37 |

在 1989 年到 2022 年期間已核准的 4,401 件發明專利，可以觀察到財團法人工業技術研

究院、中華電信股份有限公司、鴻海精密工業股份有限公司、以及本田技研工業股份有限公司都在 2003 年之前就開始在車聯網領域進行專利佈局，長期在車聯網的技術深耕。並且，由資料顯示這些企業在 2014 年之後加大投入在車聯網的技術研發。除此之外，英業達股份有限公司、聯發科技股份有限公司在 2004 年到 2014 年期間雖然較少投入車聯網的專利佈局，但卻在 2020 年到 2022 年期間加大投入的力道，被核准發明專利數較早期提升數倍，可以觀察出近幾年積極在車聯網領域進行專利佈局。

值得關注的還有近年開始大量投入在車聯網技術研發的企業，包含ＯＰＰＯ廣東移動通信有限公司、高通公司、新力股份有限公司、以及北京市商湯科技開發有限公司。從熱點來看，可以觀察到這些企業主要在 2020 年到 2022 年期間大量投入在車聯網的專利佈局，顯示出許多國家/地區積極發展在車聯網相關技術和佈局。

二、 主要技術領導企業的主要技術和應用領域情況

為分析主要技術領導企業的主要技術和應用領域情況，本研究從第一申請人和第一國際專利分類號兩個變量來分析其關聯，並且以核准發明專利數前十名的前十名第一申請人和第一國際專利分類號作為觀察目標進行討論，如表 6 所示。

表 6 核准發明專利數前十名第一申請人和第一 IPC 的專利數分佈情況

| 第一申請人 | H04W | H04L | G06F | H04N | H01L | G06Q | B60R | H04B | G02B | G02F |
|---|---|---|---|---|---|---|---|---|---|---|
| ＯＰＰＯ廣東移動通信有限公司 | 172 | 83 | 2 | | | | | 20 | | 1 |
| 高通公司 | 89 | 61 | 6 | 53 | 1 | | | 11 | | |
| 新力股份有限公司 | 46 | 17 | 2 | 14 | 10 | | | 8 | 4 | 2 |
| 英業達股份有限公司 | | 2 | 11 | | 2 | 3 | 3 | | 2 | |
| 財團法人工業技術研究院 | 23 | 6 | 4 | | 3 | 1 | 3 | 2 | 2 | |
| 中華電信股份有限公司 | 5 | 5 | 6 | 1 | | 16 | 4 | 3 | | |
| 鴻海精密工業股份有限公司 | 2 | | 4 | | 1 | 3 | 15 | | 5 | 8 |
| 本田技研工業股份有限公司 | | | | | | 3 | 5 | | | |
| 聯發科技股份有限公司 | 26 | 11 | | 10 | 1 | | | 3 | | |
| 北京市商湯科技開發有限公司 | | | 8 | 2 | | | | | | |

以核准發明專利數最多的第一申請人ＯＰＰＯ廣東移動通信有限公司、高通公司為例，其在車聯網的專利佈局主要在 H04W（無線通訊網路）、H04L（數位資訊之傳輸）。因為，ＯＰＰＯ廣東移動通信有限公司主要是設備製造業，並從事無線通訊網路和資料傳輸的技術開發。此外，高通公司主要是無線電通信技術和晶片的研發公司，所以著力於無線通訊網路和資料傳輸的專利佈局，並且跨足到 H04N（影像通信）領域，處於主要技術領導企業之一。

此外，以核准發明專利數較多的第一申請人新力股份有限公司、財團法人工業技術研究院、聯發科技股份有限公司為例，其在車聯網的專利佈局主要集中在 H04W（無線通訊網路）。因為，這些企業主要研發或生產終端設備或晶片，所以著重在無線通訊網路技術的專利佈局。另外，中華電信股份有限公司是資訊服務業，在車聯網領域主要研發資訊平台和提供資訊服務，所以中華電信股份有限公司的專利佈局主要著重在 G06Q（數據處理系統或方法）；而鴻海精密工業股份有限公司是電子製造業，主要生產電子零組件等，所以鴻海精密工業股份有限公司的專利佈局主要著重在 B60R（其他類不包括的車輛，車輛配件或車輛部件）。

### 三、 主要市場領導企業的主要技術和應用領域情況

為分析主要市場領導企業的主要技術和應用領域情況，本研究從第一專利權人和第一國際專利分類號兩個變量來分析其關聯，並且以核准發明專利數前十名的前十名第一專利權人和第一國際專利分類號作為觀察目標進行深入討論，如表 7 所示。其中，如果已核准發明專利沒有做專利轉讓或授權的情況下，第一申請人和第一專利權人的核准發明專利數情況應該是類似的，如：ＯＰＰＯ廣東移動通信有限公司、英業達股份有限公司、中華電信股份有限公司、以及本田技研工業股份有限公司等，其發明專利主要用來保護自己研發的產品。而也有部分企業做少數已核准發明專利的專利轉讓或授權，所以核准發明專利數略有變化，如：高通公司、新力股份有限公司、財團法人工業技術研究院、鴻海精密工業股份有限公司等。在本節中將對第一申請人和第一專利權人變化較大的企業進行分析。

以核准發明專利數變化較大的第一專利權人大唐移動通信設備有限公司為例，可以觀察到其在車聯網的核准發明專利數從 36 件增加到 98 件，並且其核准發明專利主要來自其母公司電信科學技術研究院有限公司。在申請時，電信科學技術研究院有限公司申請並核准較多的發明專利，在核准後再轉讓或授權給大唐移動通信設備有限公司。大唐移動通信設備有限公司和電信科學技術研究院有限公司在專利佈局的主要技術和應用領域上一致，主要從事無線通訊網路和行動設的開發，所以主要佈局在 H04W（無線通訊網路）、H04L（數位資訊之傳輸）。

此外，開曼群島商創新先進技術有限公司和阿里巴巴集團服務有限公司之間也有類似的作法，由母公司阿里巴巴集團服務有限公司申請和取得核准專利後，再把已核准發明專利轉讓或授權給開曼群島商創新先進技術有限公司；因此，開曼群島商創新先進技術有限公司以第一申請人和第一專利權人的核准發明專利數有顯著差異。並且，開曼群島商創新先進技術有限公司和阿里巴巴集團服務有限公司主要開發資訊平台提供資訊服務，所以主要佈局在 G06F（電子數位資料處理）、G06Q（數據處理系統或方法）。

表 7　核准發明專利數前十名第一專利權人和第一 IPC 的專利數分佈情況

| 第一專利權人 | H04W | H04L | G06F | H04N | H01L | G06Q | B60R | H04B | G02B | G02F |
|---|---|---|---|---|---|---|---|---|---|---|
| ＯＰＰＯ廣東移動通信有限公司 | 172 | 83 | 2 | | | | | 20 | | 1 |
| 高通公司 | 89 | 61 | 7 | 53 | 1 | | | 12 | | |
| 新力股份有限公司 | 46 | 17 | 2 | 14 | 10 | | | 8 | 4 | |
| 英業達股份有限公司 | | 2 | 11 | | 2 | 3 | 3 | | 2 | |
| 大唐移動通信設備有限公司 | 69 | 19 | | | | | | 7 | | |
| 財團法人工業技術研究院 | 22 | 6 | 3 | | 3 | 1 | 3 | 2 | 2 | |
| 中華電信股份有限公司 | 5 | 5 | 6 | 1 | | 16 | 4 | 3 | | |
| 本田技研工業股份有限公司 | | | | | | 3 | 5 | | | |
| 開曼群島商創新先進技術有限公司 | 3 | 5 | 17 | | | 18 | | | | |
| 鴻海精密工業股份有限公司 | 2 | | 4 | | 1 | 3 | 11 | | 5 | 8 |

# 伍、 結論與建議

　　有鑑於近年各國開始規範車聯網的標準，預期將帶動車聯網相關系統和設備的標準化和大量生產，帶動各個企業積極投入車聯網的專利佈局，本研究通過專利資訊檢索系統採集 1989 年到 2022 年期間已核准的 4,401 件發明專利，並且對車聯網的熱點發展趨勢、主要技術領導企業、主要市場領導企業、以及主要技術和應用領域進行深入討論與分析。由統計資料顯示，在車聯網領域的主要技術領導企業和主要市場領導企業是ＯＰＰＯ廣東移動通信有限公司和高通公司，其在車聯網領域，主要著重在無線通訊網路技術和資料傳輸，並且開終端設備和晶片等。另外，主要的技術領導企業還有新力股份有限公司、英業達股份有限公司、財團法人工業技術研究院、中華電信股份有限公司、鴻海精密工業股份有限公司、本田技研工業股份有限公司、聯發科技股份有限公司、北京市商湯科技開發有限公司等，資料顯示這些企業持續加大在車聯網的專利佈局。值得關注的還有大唐移動通信設備有限公司和開曼群島商創新先進技術有限公司，由於這些企業的已核准專利大部分通過專利轉讓或授權的方式，由其各別的母公司電信科學技術研究院有限公司和阿里巴巴集團服務有限公司提供已核准發明專利給大唐移動通信設備有限公司和開曼群島商創新先進技術有限公司，並躍居核准發明專利數前十名第一專利權人（即本研究定義的主要市場領導企業），反應出許多國家/地區的企業

積極投入車聯網的專利佈局，後續值得深入追蹤分析。

# 參考文獻